\documentstyle[12pt,epsfig]{article}
\begin{document}

\begin{center}
{\large{\bf GLUON DOMINANCE MODEL\\
 AND MULTIPARTICLE PRODUCTION}}

\vspace*{5mm}

\underline{E.S.Kokoulina}\footnote{E--mail: {\tt
kokoulin@sunse.jinr.ru}}

\vspace*{3mm}

{\it  JINR, 141980, Dubna, Moscow region, Russia}
\end{center}

\vspace*{5mm}

{\small{ \centerline{\bf Abstract} Gluon dominance model is
proposed to study multiparticle production. This model describes
multiplicity distributions in $e^+e^-$, $p\bar p$ annihilation and
$pp$ interactions. \vspace*{3mm}

{\bf 1.~~Introduction}

\vspace*{3mm}

 To investigate multiparticle
production (MP) at high energy two stage model was built  [1-4].
It is based on the use of QCD and the phenomenological scheme of
hadronisation. The model describes well multiplicity distributions
(MD) and their moments in $e^+e^-$ annihilation, $pp$ and $p\bar
p$ interactions. It confirms of fragmentation mechanism of
hadronisation in $e^+e^-$annihilation and the transition to
recombination one in hadron and nucleus interactions. It can
explain the shoulder structure in MD at higher energies and the
behavior of $f_2$ in pure $p \bar p$ annihilation at few tens
GeV/c by the inclusion of intermediate quark topologies. The
mechanism of the soft photons (SP) production as a sign of
hadronisation and estimates the emission region size of them is
proposed [5].

 \vspace*{3mm}

 {\bf 2.~~Study of $e^+e^-$
--annihilation}

\vspace*{3mm}

We began our study from $e^+e^-$ annihilation. It can be realized
through the formation of virtual $\gamma$ or $Z^0$--boson which
then decays into two quarks:
$e^+e^-\rightarrow(Z^0/\gamma)\rightarrow q\bar q.$ The $e^+e^-$--
reaction is simple for analysis, as the produced state is pure
$q\overline q$.

The perturbative QCD (pQCD) may be applied to describe the process
of parton (quark or gluon) fission at big virtuality. This stage
can be named as the cascade stage. When partons reach small
virtuality, they change into hadrons, which we observe. At this
stage we can not apply pQCD. Therefore phenomenological models are
used to describe hadronisation.

At studying MP at high energy we used ideas of A.~Giovannini [6]
to describe  quark-gluon jets as Markov branching processes. Three
elementary processes contribute into QCD jets: (1) gluon fission;
(2) quark bremsstrahlung and (3) quark pair production.
A.Giovannini had constructed a system of differential equations
for generating functions (GF) of parton ($q$, $g$) jet and
obtained MD of partons formed from quark
\begin{equation}
\label{4} P_{0}^P(Y)=\left (\frac{k_p}{k_p+\overline m}\right
)^{k_p}, \quad P_{m}^P(Y)=\frac{k_p(k_p+1)\dots(k_p+
m-1)}{m!}\left(\frac{\overline m} {\overline
m+k_p}\right)^{m}\left( \frac{k_p}{k_p+\overline m}\right)^{k_p},
\end{equation}
where $\overline m$ is the mean multiplicity, $k_p$-- parameter.
These MD are known as negative binomial distributions (NBD). The
GF for them is
\begin{equation}
\label{6} Q^{(q)}(z,Y)=\sum\limits_{m=0}^{\infty} z^{m} P_{m}(Y)=
 \left[1+ \overline m/k_p (1-z)\right]^{-k_p}.
\end{equation}

Two stage model [1-2] has taken (\ref{4}) to describe the cascade
stage and adds to it a sub narrow binomial distribution (BD) for
the hadronisation stage. We had chosen BD basing on the analysis
of experimental data in $e^+e^-$- annihilation lower than 9 GeV.
Second correlation moments were negative at these energyies. We
add the hadronisation stage to the parton one with the help of a
factorization. MD in this process can be written as follows:
\begin{equation}
\label{7} P_n(s)=\sum\limits_mP^P_mP_n^H(m,s),
\end{equation}
where $P_m^P$ is MD for partons (\ref{4}), $P_n^H(m,s)$ - MD for
hadrons produced from $m$ partons at the stage of hadronisation.
Further we substitute variable $Y$ on a center of masses energy
$\sqrt s$. MD of hadrons $P_n^H$ formed from one parton and their
GF $Q^H_p(z)$ are [1-2]
\begin{equation}
\label{12} P_n^H=C^n_{N_p}\left(\frac{\overline n^h_p}
{N_p}\right)^n\left(1-\frac{\overline n_p^h}
{N_p}\right)^{N_p-n},\quad Q^H_p=\left[1+\frac{\overline n^h_p}
{N_p}(z-1)\right]^{N_p},
\end{equation}
where $C_{N_p}^n$ - binomial coefficient, $\overline n^h_p$ and
$N_p$ ($p=q,g$) have a sense of mean multiplicity and maximum of
secondary hadrons are formed from parton at the stage of
hadronisation.

 MD of hadrons in $e^+e^-$ annihilation are
determined by convolution of two stages (cascade and
hadronisation)
\begin{equation}
\label{13} P_n(s)=\sum\limits_{m=o}^{\infty}
P_m^P\frac{1}{n!}\frac{\partial^n}{\partial z^n}
(Q^H)^{2+m}|_{z=0},
\end{equation}
where $2+m$ is the total number of partons (two quarks and $m$
gluons).

The parameter $\alpha=N_g/N_q$ distinguishes the hadrons produced
from quark or gluon. Introducing expressions (\ref{4}), (\ref{12})
in (\ref{13}) we obtain MD of hadrons in $e^+e^-$ annihilation
($N=N_q$, $\overline n^h=\overline n^h_q$)

\begin{equation}
\label{15} P_n(s)= \sum\limits_{m=0} ^{M_g}P_m^PC^n_{(2+\alpha
m)N} \left(\frac{\overline n^h} {N}\right)^n\left(1-\frac
{\overline n^h}{N}\right)^{(2 +\alpha m)N-n}.
\end{equation}

The results of comparison of expression (\ref{15}) with
experimental data [7] are led in [2]. The significance of $\alpha
$ is equal to $0.2$ with some deviations. If we know from our
comparison $\alpha , N_q$ and $\overline n^h_q$  then we can
determine for gluon $N_g=\alpha N$ and $\overline n_g^h= \alpha
\overline n^h$. It is surprising these parameters remain constant
without considerable deviations in spite of the indirect finding:
$N_g\sim 3-4$ and $\overline n_g^h\sim 1$ (Fig. 1). This behaviour
is evidence of the universality of gluon hadronisation. As $\alpha
< 0.2$ we come to a conclusion that hadronisation of gluons is
softer than of quarks.

\begin{figure}[h!]
 \leavevmode
\begin{minipage}[b]{.475\linewidth}
\centering
\includegraphics[width=\linewidth, height=2.8in, angle=0]{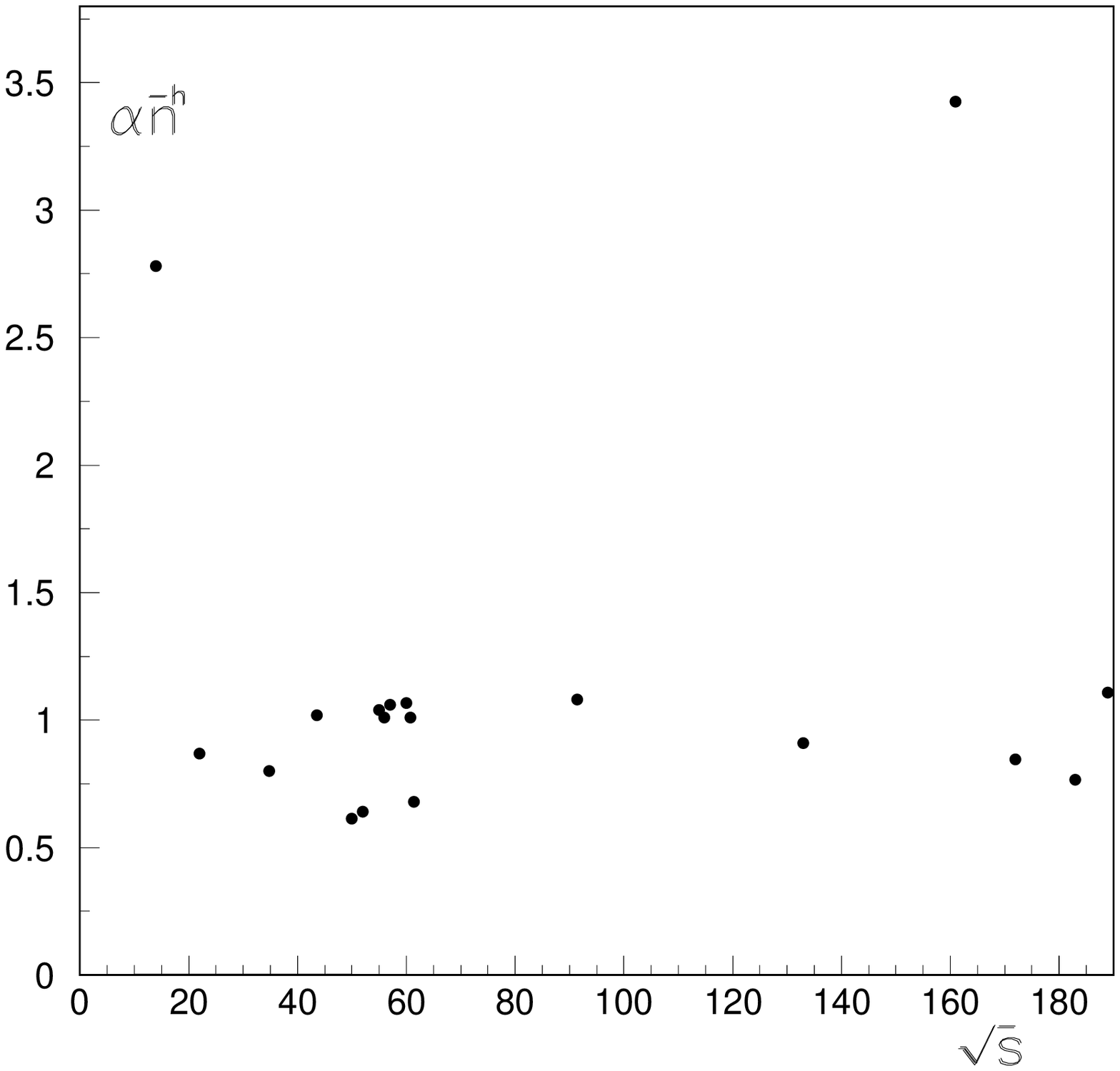}
\caption{$N_g=\alpha N_q$.} \label{secondfig}
\end{minipage}\hfill
\begin{minipage}[b]{.475\linewidth}
\centering
\includegraphics[width=\linewidth, height=3in, angle=0]{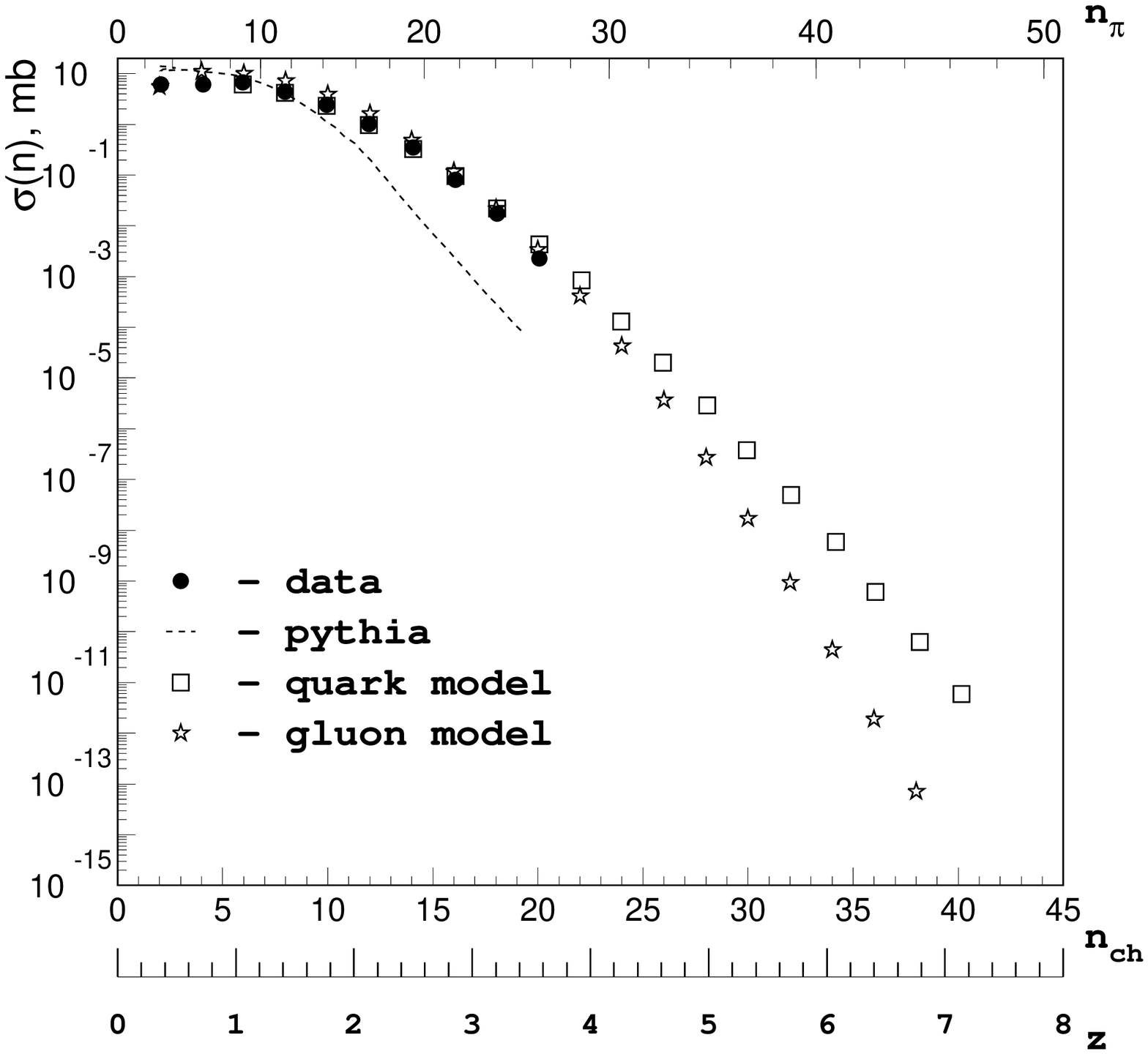}
\caption{$\sigma (n)$ in $pp$.} \label{threddfig}
\end{minipage}\\
\end{figure}

\vspace*{3mm}

 {\bf 3.~~Study of $pp$ interactions}

\vspace*{3mm}

The study of MD in $pp$ interactions is implemented in the
framework of the project "Thermalization" [8]. This project is
aimed at studying the collective behavior of secondary particles
and advancement to the high multiplicity region (HMR) beyond of
available data [9] in proton-proton interactions at 70 GeV/c.

The calculation by the MC PHYTHIA code has shown that the standard
generator predicts a value of the cross section which is in a good
agreement with the experimental data at small multiplicity
($n_{ch}<10$) but it underestimates the value $\sigma(n_{ch})$ by
two orders of the magnitude at $n_{ch}=18$ (Fig. 2). The existing
models are sensitive in HMR [10] (Fig. 2)

We are guided by a scheme building of hadron interactions to
describe MD with the quark-gluon language as well as to
investigate HMR. The existing models are sensitive in this region
[e.g. 8]. We consider that at the first stage of $pp$ interactions
the initial quarks and gluons take part in the formation of
quark-gluon system (QGS). On the second stage they can transform
to hadrons. We proposed two schemes. In the first scheme
hadroproduction is studied with account of the parton fission
inside the QGS. If we are not interested in what is going inside
QGS, then come to the second scheme [3].

When we took model where some of quarks and gluons from protons
participate in the production of hadrons the parameters of that
model differed very much from parameters obtained in $e^+e^-$-
annihilation, especially for hadronisation. That is why the scheme
with active quarks was rejected, and quarks from protons were
remained inside of the leading particles. In according of this
result all of the newly born hadrons are formed by gluons. Such
gluons we name active and a model involving theirs into
hadroproduction - the gluon dominance model (GDM) [4].

Both of schemes describe well MD at 69 GeV/c. From the comparison
with experimental data [9] the maximum number of active gluons
appeared in QGS is restricted to 6, $\overline n_g^h=1.63\pm
0.12,$ the part of evaporated gluons from QGS equal to $0.47 \pm
0.01$. We conclude from this research that the branch processes
are weak. A maximal possible number of hadrons from the gluon in
first scheme looks very much like the number of partons in the
glob of cold quark-gluon plasma of L.Van Hove [11]. After the
evaporation the part of active gluons do not convert into hadrons.
They stay in QGS and become sources of soft photons (SP) [5].

From neutral mesons data [12] hadronization parameters of $\pi ^0$
were founded (the second scheme). The analysis of the mean
multiplicity of $\pi^0$-- mesons  versus the number of charged
particles $n_{ch}$ gives the limitations to the number of neutral
mesons at given $n_{ch}$ [3]. The obtained estimations of
probabilities for a production charged and neutral hadrons from
gluon at the its passing of hadronization permits to get in
framework of GDM [4] "the charged hadron/pion" ratio in pp
interactions, which is in an agreement with data [13].

The application of GDM to describe MD in the region 102-800 GeV/c
[14] in both schemes leads to good results [5]. Parameters of GDM
in this domain are given in Table 1.
\begin{center}
Table 1. Parameters of TSTM.
\end{center}
\renewcommand{\tablename}{Table}
\begin{center}
\begin{tabular}{||c||c|c|c|c|c|c||c||}
\hline \hline
$p,$ GeV/c & $\overline m$ & $M_g$ & $N$&$\overline n^h$&$\Omega $&$\chi^2/$ndf\\
\hline \hline
102  & $2.75\pm 0.08$ & 8  & $3.13\pm 0.56$ & $1.64 \pm 0.04$ & $1.92\pm 0.08$ & 2.2/5\\
\hline
205  & $2.82 \pm 0.20$ & 8 & $4.50 \pm 0.10$& $2.02 \pm 0.12$& $2.00 \pm 0.07$  &2.0/8  \\
\hline
300& $2.94 \pm 0.34$ & 10  &$4.07 \pm 0.86$ & $2.22 \pm 0.23$& $1.97 \pm 0.05$  & 9.8/9 \\
\hline
405& $2.70 \pm 0.30$& 9& $4.60 \pm 0.24$ & $2.66 \pm 0.22$ & $1.98 \pm 0.07$ &16.4/12 \\
\hline
800  &  $3.41 \pm 2.55$& 10&$ 20.30 \pm 10.40 $& $2.41 \pm 1.69$ & $2.01 \pm 0.08$&10.8/12\\
\hline \hline
\end{tabular}
\end{center}
A growth of $\overline n^h_g$ in $pp$ interactions indicates a
change mechanism of hadronization of gluons in comparison with
$e^+e^-$ annihilation. It is considered that in the $e^+e^-$
process the partons transform to hadrons by the fragmentation
mechanism (the thermal medium is absent). Our analysis gives
$\overline n_g^h \sim 1$ [2]. The recombination is specific for
the hadron interactions. In these processes pairs from gluons
appear almost simultaneously and recombine to various hadrons
[15]. The value $\overline n_g^h$ becomes bigger $\sim 2-3$, that
indicates this transition.

At the top energy (200-900 GeV) the shoulder structure appears in
$P_n$ [16]. The comparison of data with one NBD does not describe
data. But the weighted superposition of two NBD gives a good
description of the shoulder structure $P_n(s)$ [6]. We modify our
GDM considering that the gluon fission is realized at higher
energies. The independent evaporation of gluons sources of hadrons
occures by single gluons and also groups from two and more fission
gluons. We name such groups of gluons - clans, too. Their
independent emergence and following hadronization is a content of
GDM. MD in GDM with two kinds of clans are:
$$ P_n(s)=\alpha _1\sum\limits_{m_1=0}^{Mg_1}\frac{e^{-\overline
m_1} \overline m_1^{m_1}}{m_1!} C^{n-2}_{m_1\cdot
N}\left(\frac{\overline n^h}
{N}\right)^{n-2}\left(1-\frac{\overline n^h} {N}\right)^{m_1\cdot
N-(n-2)}+
$$
\begin{equation}
\label{48} + \alpha
_2\sum\limits_{m_2=0}^{Mg_2}\frac{e^{-\overline m_2} \overline
m_2^{m_2}}{m_2!} C^{n-2}_{2\cdot m_2\cdot N}\left(\frac{\overline
n^h} {N}\right)^{n-2}\left(1-\frac{\overline n^h}
{N}\right)^{2\cdot m_2\cdot N-(n-2)},
\end{equation}
where $\alpha _1$ and $\alpha _2$ are the contribution single and
double gluon clans ($\alpha _1 + \alpha _2 =1$). The comparison
(\ref{48}) with experimental data for proton interactions at
$\sqrt s= 62.2$ GeV [16] gives the following values of parameters:
$N=7.06\pm 3.48$, $\overline m_1 =3.59 \pm 0.03$, $\overline
m_2=1.15\pm 0.25$, $\overline n_h=3.23\pm 0.14$, $Mg_1=8$,
$Mg_2=4$, $\alpha _1/\alpha _2 \sim 1.8$ at $\chi
^2/$ndf=$9.12/13$ (Fig. 3).

The main feature of our GDM approach is the dominance of active
gluons in MP. We expect the emergence of them in experiments at
high energy (SPS, RHIC) and the formation of quark-gluon plasma.
We consider that our QGS can be a candidate for this.

\vspace*{3mm}

 {\bf 3.~~Study of $p\bar p$ annihilation}

\vspace*{3mm}

In the midst of hadron interactions the $p\bar p$ annihilation
shows up especially interesting. Experimental data at tens GeV/c
[18] point out on some maxima in differences between $p\bar p$ and
$pp$ inelastic topological cross sections what may witness about
the contribution of different mechanisms of MP
\begin{equation}
\label{50} \Delta \sigma _n(p\overline p -pp)= \sigma _n(p
\overline p) - \sigma _n (pp).
\end{equation}
The important information about them may be picked out from the
analysis of the second correlative moment for negative charged
particles $f_2^{--}$
\begin{equation}
\label{49} f_2^{--}=\overline {n_-(n_--1)}-{\overline{n}_-}^2.
\end{equation}
The negative value of second correlative moments is characteristic
for a more narrow MD than Poisson, and they indicate the
predominance of the hadronization stage in MP.

At the initial stage of annihilation three valent $q\overline
q$-pairs ($uud$ and $\overline u \overline u \overline d$) are.
They can turn to the "leading" mesons which consist from: a)
valent quarks or b) valent and vacuum quarks. In the case a) only
three "leading" neutral pions (the "0" topology) or two charged
and one neutral "leading" mesons ("2" - topology) may form. In b)
case the "4"- and "6"- topology is realized for "leading" mesons.
We suggest that the formation neutron and antineutron (a charge
exchange) can be realized.

We can suggest a simple scheme when some of active gluons are
formed at the moment of annihilation and fragment into hadrons. If
the GF for a single active gluon $Q_1(z)=[1+\overline n/N(z-1)]^N$
[3], then $f_2=Q_1^{''}(z)|_{z=1} -[Q_1(z)|_{z=1}]^2 = -(\overline
n^h)^2/N < 0.$ Reciprocally for m gluons GF and $f_2$ will be
\begin{equation}
\label{53} Q_m(z)=[1+\overline n^h/N(z-1)]^{mN}, \quad
f_2=-m(\overline n^h)^2/N.
\end{equation}
We consider that if $m$ grows while increasing the energy of the
colliding particles then $f_2$ will decrease  almost linearly from
$m$. Such behavior qualitatively agrees with experimental data
[18]. According to GDM for $p\overline p$ annihilation and taking
into account three intermediate charged topologies and active
gluons, GF $Q(z)$ for final MD may be written as the convolution
gluon and hadron components:
$$Q(z)=c_0\sum\limits_m^{M_0}
P_m^G [1+\frac{\overline n^h}{N}(z-1)]^{mN}+c_2\sum\limits_m^{M_2}
z^2 P_m^G [1+\frac{\overline n^h}{N}(z-1)]^{mN}
+c_4\sum\limits_m^{M_4} z^4 P_m^G [1+\frac{\overline
n^h}{N}(z-1)]^{mN}.
$$
The parameters of $c_0$, $c_2$  and $c_4$ are determined as the
part of intermediate topology ("0", "2" or "4") to the
annihilation cross section ($c_0 + c_2 + c_4 =1).$ For the sake of
a simplicity we are limited by cut Poisson distribution with the
finite number of gluons for $P_m^G$. The comparison of the
experimental data (Fig. 4) gives the following values of
parameters: $\overline m=3.36 \pm 0.18$, $N=4.01\pm 0.61$,
$\overline n^h$=1.74 $\pm $0.26, the ratio $c_0$ : $c_2$ : $c_4$ =
15 : 40 : 0.05 at $\chi ^2/ndf=5.77/4$ and $M_0 \sim M_2 \sim
1-2$, $M_4\sim 4$.

\begin{figure}[h!]
 \leavevmode
\begin{minipage}[b]{.475\linewidth}
\centering
\includegraphics[width=\linewidth, height=3in, angle=0]{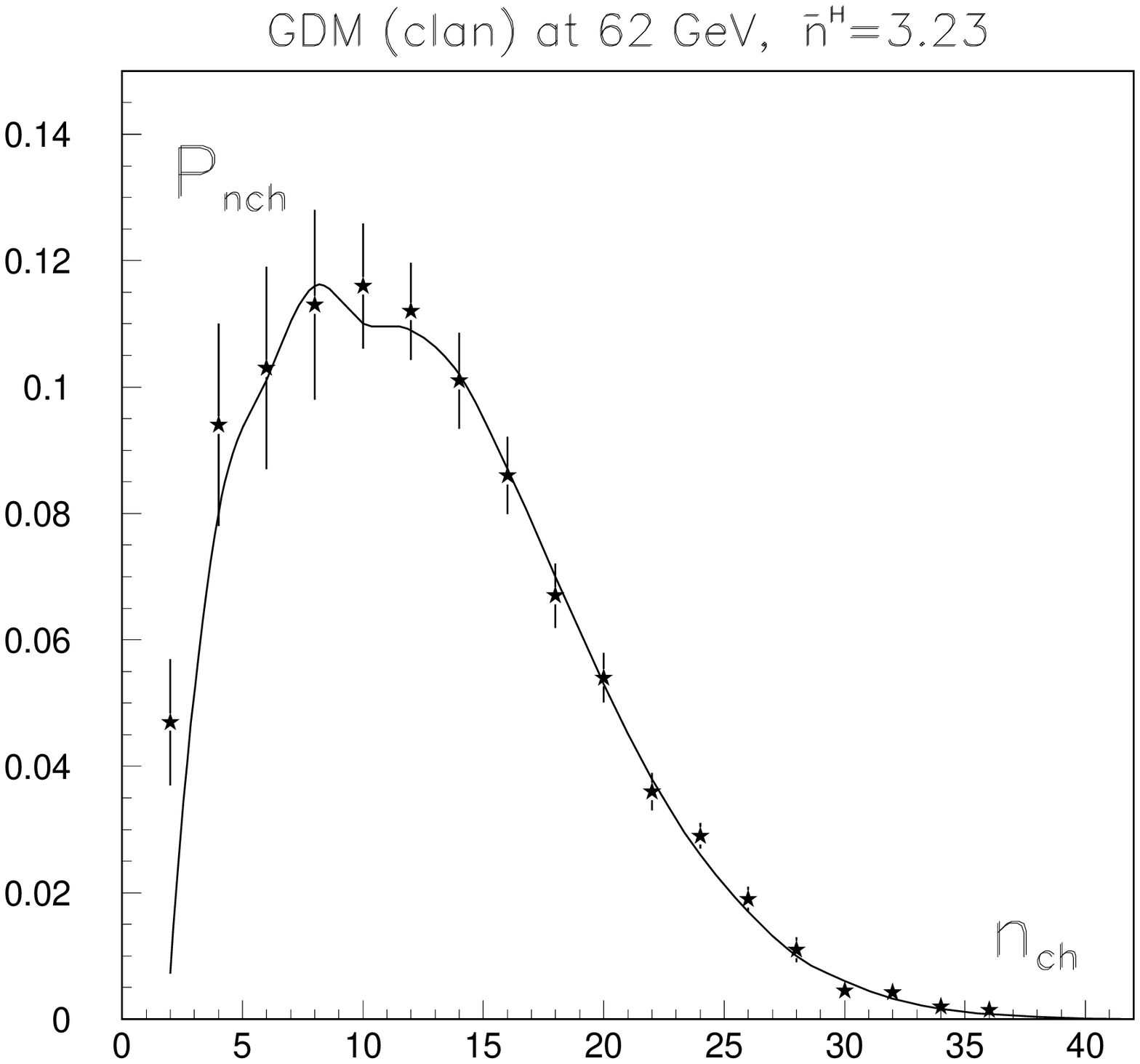}
\caption{MD in GDM (clan).} \label{secondfig}
\end{minipage}\hfill
\begin{minipage}[b]{.475\linewidth}
\centering
\includegraphics[width=\linewidth, height=3in, angle=0]{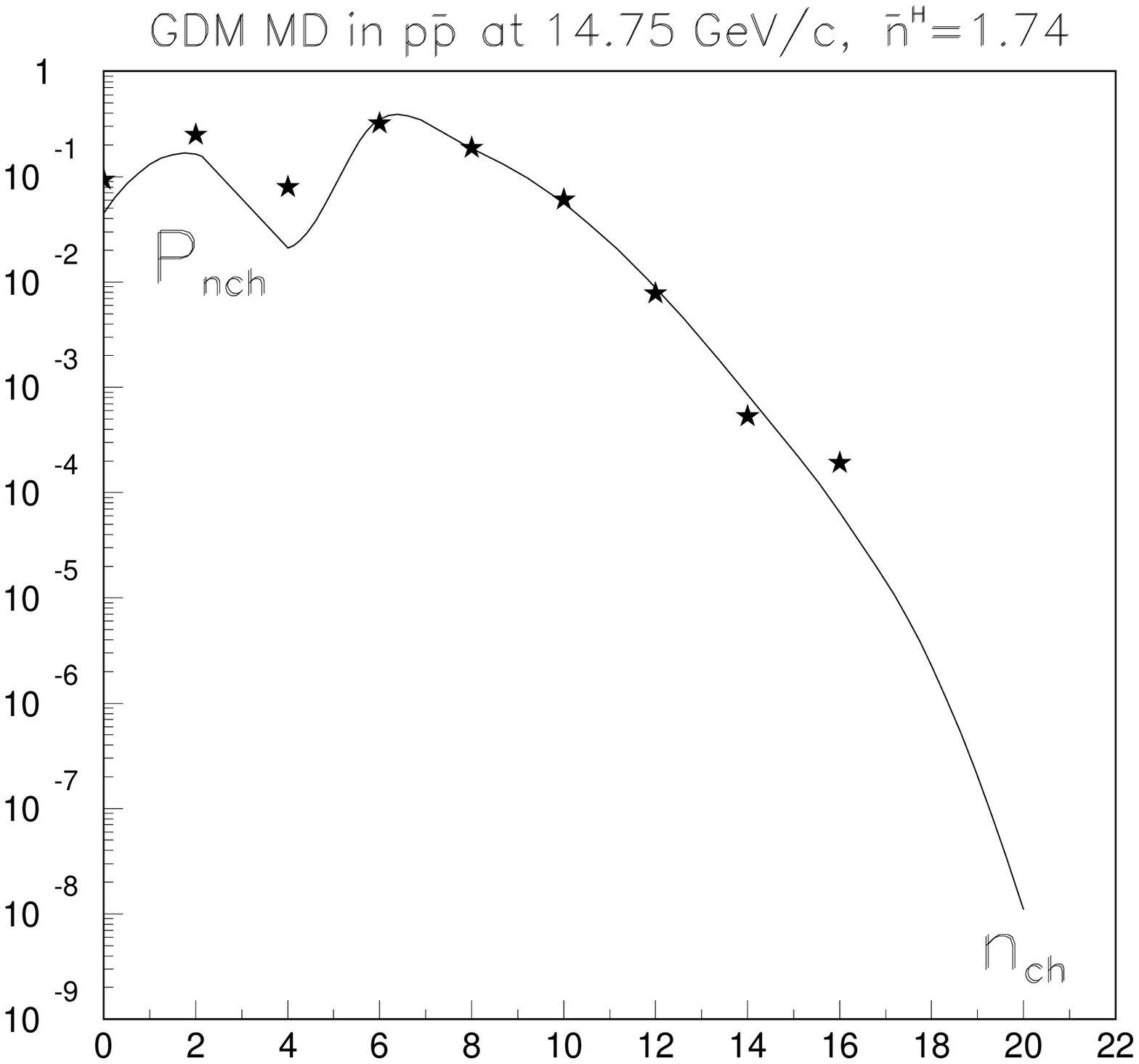}
\caption{MD in $p\overline p$ at 14.75 GeV/c.} \label{threddfig}
\end{minipage}\\
\end{figure}

\vspace*{3mm}

 {\bf 3.~~Soft photons}

\vspace*{3mm}

The production of photons in particle collisions at high energies
was studied in many experiments [19]. In project "Thermalization"
it is planned to investigate low energetic photons with $p_t \leq
0.1GeV/c$ and $x \leq 0.01$ [8]. These photons are named soft
photons (SP). Experiments shown that measured cross sections of SP
are several times larger than the expected ones from QED inner
bremsstrahlung. Phenomenological models were proposed to explain
the SP excess (the glob model of Lichard and Van Hove [11]).

We consider that at a certain moment the QGS or excited new
hadrons may set in an almost equilibrium state during a short
time. That is why, to describe massless photons, we have used the
black body emission spectrum. At 70 GeV/c an inelastic cross
section is equal to $\sim 40 mb$, the SP formation cross section
is about $4 mb$ [8] and since $ \sigma _{\gamma} \simeq n_{\gamma
}(T)\cdot \sigma _{in}$ then $n_{\gamma }\approx 0.1$. If
$n_{\gamma }$ and temperature $T(p)$ ($p$--momentum) are known,
then we can estimate the emission region size $L$ of SP. The
obtained values $L\sim 4-6 fm$ [4,5] that is reasonable size.

In our study we had undertaken an attempt to give description in
different processes of MP by means of a unified approach based on
quark-gluon picture with using the phenomenological hadronisation.
We have obtained agreements of our schemes with experimental data
in $e^+e^-$, $p\overline p$ annihilation and $pp$ and nucleus
collisions in a very wide energy domain.

\vspace*{3mm}

{\bf References}

\vspace*{3mm}

\begin{tabular}{rp{135mm}}

 [1] &V.I.~Kuvshinov and E.S.~Kokoulina, Acta Phys.Polon., {\bf  B13},
p.533, 1982.\\

 [2] &E.S.~Kokoulina, XI Ann.Sem. NPCS, Minsk, Belarus, 2002,
hep-ph/0209334; E.S.Kokoulina, ISMD32, 2002.\\

 [3] &E.S.~Kokoulina and V.A.~Nikitin, hep-ph/0308139;
E.S.Kokoulina, Acta Phys.Polon., {\bf B35}, p.295, 2004.\\

 [4] &E.S. Kokoulina and V.A. Nikitin, ISHEPP, 2004,
hep-ph/0502224; P.F. Ermolov et al., ISHEPP, 2004,
hep-ph/0503254.\\

 [5] &M.K.Volkov et al., Particles and Nuclei, Letters,
{bf 1}, p.16, 2004.\\

 [6] &A.~Giovannini, Nucl.Phys., {\bf B 161}, p.429, 1979;
A.~Giovannini and R.~Ugocioni, hep-ph/0405251.\\

 [7] &W.Braunschweig et al. Z. Physik $\bf {C 45}$, 193
(1989); H.W.Zheng et al. Phys. Rev. $\bf {D 42}$, 737 (1990);
P.Abreu et al. Z.Phys. $\bf {C 52}$, 271 (1991); P.D.Acton et. al.
Z.Physik. $\bf {C53}$, 539 (1992), Alexander G. et al. Z.Physik.
$\bf {C 72}$, 191 (1996); K.Acketstaff et al. Z.Physik. $\bf {C
75}$, 193 (1997); G.Abbiendi et al. CERN-EP/99-178.\\

 [8] &P.F.Ermolov et al. Yad. Phys., {\bf 67}, p.108, 2004.\\

 [9] &V.V.Babintsev et al. IHEP preprint M-25, Protvino (1976); V.V.
Ammosov et al., Phys. Let., {\bf 42B}, p.519, 1972.\\

 [10] &O.G. Chikilev and P.V. Chliapnikov, Yad. Phys.,{ \bf55 },
 p.820, 1992.\\

 [11] &P.Lichard and L.Van Hove, Phys.Let., {\bf B245}, p.605,
 1990.\\

 [12] &V.S. Murzin and L.I. Sarycheva.
\emph{Interactions of high energy hadrons}. Nauka, Moscow, 1983,
pp.194-207; V.G. Grishin, Phys. El.Part. and At.Yad., {bf 10},
p.608, 1979; K. Jaeger et al.,Phys. Rev., {bf D11}, p.2405,
1975.\\

 [13] &K. Adcox et. al, Nucl.Phys., {bf A757}, p.184, 2005;
B. Alper et al, Nucl. Phys., {bf B100}, p.237, 1975.\\

 [14] &C.Bromberg et al., Phys. Rev. Lett. {bfseries 31}, 1563 (1973);
W.M.Morse et al., Phys.R ev., {\bf D15}, 66 (1977); S.Barish et
al., Phys. Rev., {\bf D9}, 2689 (1974); A.Firestoun et al., Phys.
Rev.,{\bf D10}, 2080 (1974); P.Slattery, Phys.Rev., {\bf D7}, 2073
(1973).\\

 [15] &R.C.~ Hwa and C.B.~ Yang, Phys. Rev., {bf C67}, p.034902,
2003; R.Hwa.\\

 [16] &G.I. Alner et al., Z. Phys., {bf C33}, p.1, 1986;
R.E.~ Ansorge et al., Z.Phys., {\bf C43}, p.357, 1989.\\

 [17] &A.~ Breakstoune et al., Phys.Rev., {bf D30}, p.528,
1984.\\

 [18] &J.G. Rushbrooke and B.R. Webber, Phys. Rep., {bf C44},
p.1, 1978.\\

 [19] &P.V.Chliapnikov et al., Phys. Let., {\bf 141B}, 276 (1984);
S.Banerjee et al., Phys.Let.,{\bf B305},182 (1983); J.Antos et
al., Z.Phys., {\bf C59}, 547 (1993); J.F.Owens., Rev. Mod.Phys.,
{\bf 59},465 (1987).\\

\end{tabular}

\end{document}